\documentclass[twocolumn,prl,secnumarabic,superscriptaddress]{revtex4}
\usepackage{amsfonts}
\usepackage{amsmath}
\usepackage{amssymb}
\usepackage{graphicx}
\usepackage{epstopdf}

\newcommand{\Pinjhat}{\hat{P}_{\mathrm{inj}}}
\newcommand{\Pdethat}{\hat{P}_{\mathrm{det}}}
\newcommand{\Pinj}{P_{\mathrm{inj}}}
\newcommand{\Pdet}{P_{\mathrm{det}}}
\newcommand{\sdu}{\sigma_{\mathrm{D}\uparrow}}
\newcommand{\sdd}{\sigma_{\mathrm{D}\downarrow}}
\newcommand{\sru}{\sigma_{\mathrm{R}\uparrow}}
\newcommand{\srd}{\sigma_{\mathrm{R}\downarrow}}
\newcommand{\sigu}{\sigma_{\uparrow}}
\newcommand{\sigd}{\sigma_{\downarrow}}
\newcommand{\Vl}{V_{\mathrm{g}}^{\mathrm{D}}}
\newcommand{\Vr}{V_{\mathrm{g}}^{\mathrm{R}}}
\newcommand{\Vinj}{V_{\mathrm{g}}^{\mathrm{inj}}}
\newcommand{\Vdet}{V_{\mathrm{g}}^{\mathrm{det}}}
\newcommand{\Vnl}{V_{\mathrm{nl}}}
\newcommand{\vnl}{V_{\mathrm{nl}}}

\begin{document}

\title{Measurement of spin-dependent conductivities in a two-dimensional electron gas}

\author{H. Ebrahimnejad}
\affiliation{Department of Physics and Astronomy, University of British Columbia,Vancouver, British Columbia V6T 1Z4, Canada}
\author{Y. Ren}
\affiliation{Department of Physics and Astronomy, University of British Columbia,Vancouver, British Columbia V6T 1Z4, Canada}
\author{S.M. Frolov}
\affiliation{Department of Physics and Astronomy, University of British Columbia,Vancouver, British Columbia V6T 1Z4, Canada}
\affiliation{Kavli Institute of Nanoscience, Delft University of Technology, 2628CJ Delft, The Netherlands}
\author{I. Adagideli}
\affiliation{Faculty of Engineering and Natural Sciences, Sabanci University, Tuzla, Istanbul 34956, Turkey}
\author{J.A. Folk}
\affiliation{Department of Physics and Astronomy, University of British Columbia,Vancouver, British Columbia V6T 1Z4, Canada}
\author{W. Wegscheider}
\affiliation{Laboratorium f\"{u}r Festk\"{o}rperphysik, ETH Z\"{u}rich, 8093 Z\"{u}rich, Switzerland}

\date{\today}

\begin{abstract}
Spin accumulation is generated by injecting an unpolarized charge current into a channel of GaAs two-dimensional electron gas subject to an in-plane magnetic field, then measured in a non-local geometry.  Unlike previous measurements that have used spin-polarized nanostructures, here the spin accumulation arises simply from the difference in bulk conductivities for spin-up and spin-down carriers.  Comparison to a diffusive model that includes spin subband splitting in magnetic field suggests a significantly enhanced electron spin susceptibility in the 2D electron gas.
\end{abstract}

%Pure spin current measurements are performed on narrow channels defined in a GaAs/AlGaAs two-dimensional electron gas subject to an in-plane magnetic field. A net spin current is observed even when the injected charge current is unpolarized, arising from the difference in bulk conductivities for spin-up and spin-down carriers. Comparison to a diffusive model that includes spin subband splitting in magnetic field suggests a significantly enhanced electron spin susceptibility in the 2D electron gas.

\maketitle

The connection between charge and spin transport in semiconductor quantum wells has significant implications for both science and technology.  From a technological point of view, future spintronic devices will depend on spin transport with charge-based readout and control.\cite{Zutic}  From a scientific point of view, the effect of spin on electrical conductivity remains a fertile area of research.  For example, it is believed that spin drives a metal-insulator transition in two-dimensional systems at low density, but the mechanism is unclear.\cite{Gao, Piot}  Conversely, electron-electron interactions lead to an enhanced spin susceptibility, but the origin and magnitude of the effect is still debated.\cite{Tutuc2, Gold, Liang, Tan}

These questions are often addressed experimentally by measuring the change in electrical conductivity as carriers are polarized using an in-plane magnetic field.   Despite the simplicity of such a measurement, however, the results are not easy to interpret.  Data are difficult to match with theoretical predictions, in part because spin and orbital effects of the in-plane field are hard to separate, and because theoretical analysis is not yet well-developed for remotely doped structures such as GaAs/AlGaAs quantum wells.\cite{Piot, Gold}  Clearer experimental insights may be gained from measurements that distinguish spin transport from charge transport.\cite{Johnson, Jedema, PotokPRL02, Frolov}

In this Letter, we use a spin-sensitive measurement to quantify the difference between spin-up and spin-down conductivities, $\sigma_\uparrow$ and $\sigma_\downarrow$, in a GaAs/AlGaAs 2DEG subject to an in-plane magnetic field.  Charge currents are injected into a narrow 2DEG channel using a quantum point contact (QPC) on the 2$e^2/h$ plateau.  A spin accumulation is developed in the channel when $\sigma_\uparrow\neq\sigma_\downarrow$, even if the injected current is strictly unpolarized.    Spin accumulation in nonmagnetic 2DEGs has previously been generated using the polarized current resulting from transport through a spin-selective nanostructure such as a quantum dot or a quantum point contact.\cite{PotokPRL02, PotokPRL03, HansonPRB04}  Our measurements show that the accumulation due to an unpolarized current can be nearly as large as that due to fully-polarized charge current, but with opposite sign.  The magnitude of the effect provides an estimate of the spin susceptibility in the channel.

\begin{figure}
  \includegraphics[scale=1]{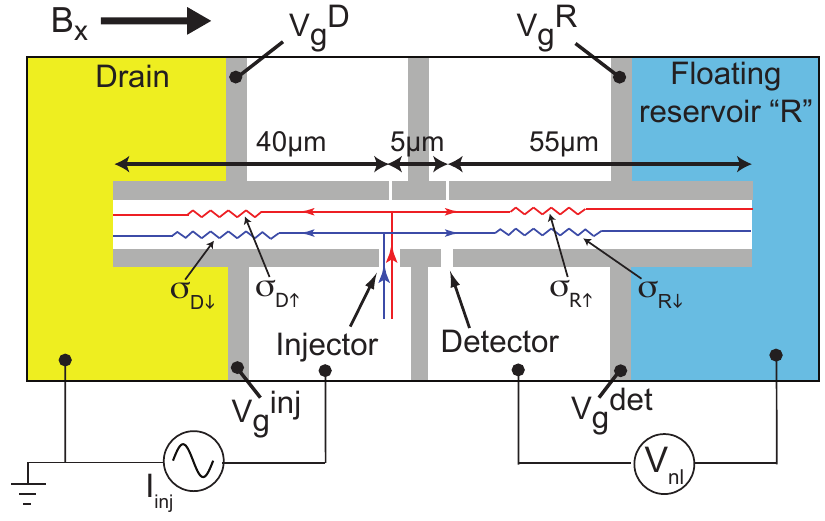}
  \caption{\textbf{(a)} Device schematic showing gates (grey) that control injector and detector QPCs, $\Vinj$ and $\Vdet$, which are separated by 5$\mu m$ ``middle" segment,  and that control the densities of the drain and reservoir sides of the channel, $\Vl$ and $\Vr$. Circuits for spin-up and spin-down currents are indicated by $\sigma_{D\uparrow}$,$\sigma_{R\uparrow}$ and $\sigma_{D\downarrow}$,$\sigma_{R\downarrow}$.  Gates are depleted even for small positive applied voltages due to bias cooling.}
 \label{device}
\end{figure}

Measurements were performed on channels with widths of one and two microns, defined electrostatically in a high-mobility GaAs/AlGaAs 2DEG (bulk electron density $n_{s}=1.11\times10^{11}$cm$^{-2}$ and mobility $\mu=4.44\times10^{6}\mathrm{cm}^{2}/\mathrm{Vs}$ measured at $T=1.5$K).  This paper contains data from two of the channels.  The channels were aligned along $\hat{x}$, defined as the [110] GaAs crystal axis (Fig.1).  A charge current, I=2 nA, injected midway along the channel through the injector QPC at $x=0$, was drained on the left end of the channel. The reservoir at the right end was floating, so charge current could flow only to the left of the injector.  A detector QPC to the right of the injector at $x_{det}=5\mu m$ served as a nonlocal voltage probe.\cite{Johnson, Frolov}  Data were taken at T=300mK in in-plane magnetic fields up to 11T.

The polarizations of the injector and the detector contacts could be tuned by gate voltages $\Vinj$ and $\Vdet$. QPC conductance at low temperature and high magnetic field is quantized in units of $Ne^2/h$, where $N=N_{\uparrow}+N_{\downarrow}$ is the total number of spin resolved subbands.\cite{Wees, Wharam} The spin-up and spin-down subbands are added sequentially, so the polarization of QPC transmission $P\equiv(N_{\uparrow}-N_{\downarrow})/(N_{\uparrow}+N_{\downarrow})=1, 0, 1/3, 0, 1/5$ for N=1, 2, 3, 4, 5. The injected current is polarized when $N_{inj}$=1, 3, 5, etc. For $P_{det}$=1 the potential of the detector adjusts to the spin-up chemical potential $\mu_\uparrow$ in the channel at $x=x_{det}$. For $P_{det}=0$ the detector adjusts to the average chemical potential $\mu_{av}\equiv(\mu_{\uparrow}+\mu_{\downarrow})/2$.  The nonlocal voltage, $\vnl$, reflects the difference between the detector potential and the potential in the floating reservoir $\mu=\mu(R)$, assumed to be at equilibrium.

The two panels of Fig. 2(a) show the spatial dependence of spin chemical potentials that might be expected assuming spin-independent conductivity and neglecting spin relaxation. For $P_{inj}=0$, $\mu_{\uparrow}=\mu_{\downarrow}$ in the entire channel. For $P_{inj}=1$, a non-equilibrium accumulation of spin-up carriers builds up in the channel near the injector, leading to $\mu_{\uparrow}>\mu_{\downarrow}$.  Spins diffuse left and right to the reservoirs, which are assumed to be at equilibrium ($\mu_{\uparrow}=\mu_{\downarrow}$).  The chemical potential of the (floating) right reservoir, $\mu(R)$, equilibrates midway between $\mu_{\uparrow}$ and $\mu_{\downarrow}$ at the injector to satisfy the condition of zero net current. One might therefore expect a positive nonlocal voltage $V_{nl}=\mu_\uparrow(x_{det})-\mu(R)$ for \{$P_{inj},P_{det}$\}=\{1,1\} and a zero voltage for \{0,1\} ($V_{nl}=\mu_\uparrow(x_{det})-\mu(R)$) as well as for \{1,0\} ($V_{nl}=\mu_{av}(x_{det})-\mu(R)$).

\begin{figure}
  \includegraphics[scale=1]{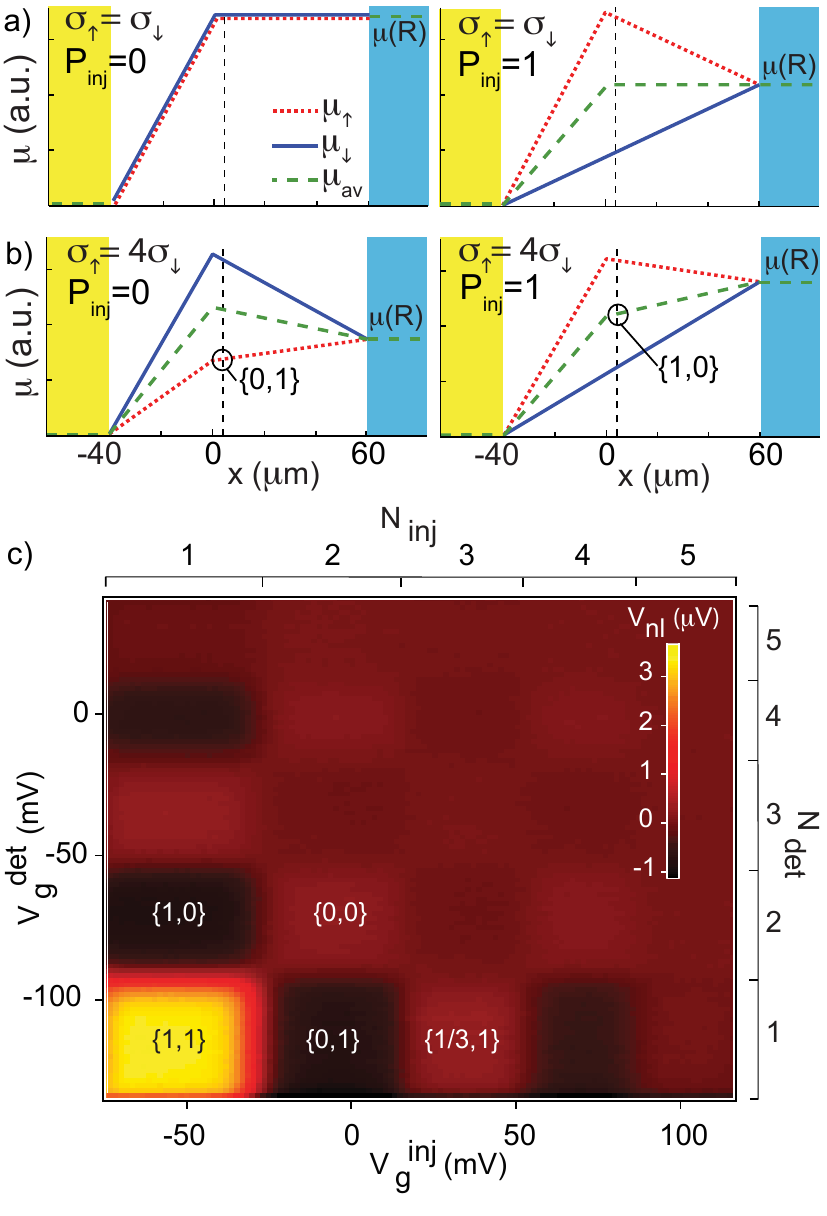}
  \caption{\textbf{(a,b)}  Spatial dependence of spin-up and spin-down chemical potentials. The injector is at x=0, dashed lines mark the detector position $\rm{x=x_{det}}$, where $\rm{x_{det}}=5\mu m$ in these schematics. Circles in (b) show chemical potentials at the detector that are below the reservoir potential, giving rise to negative signals for \{0,1\} and \{1,0\}. \textbf{(c)} Measured nonlocal voltage for $B_x=10.5$T.}
  \label{negsignal}
\end{figure}

Figure 2(c) shows a typical measurement of $\vnl$. The positive $\vnl$ resulting from a fully polarized injector and detector is clearly visible in the \{1,1\} region. This signal was investigated in detail in Ref.~\onlinecite{Frolov}. Contrary to the simple picture presented in Fig.~2(a), however, the nonlocal signal is not zero for \{1,0\} or \{0,1\}---in fact, it is negative.  Similar negative voltages were observed whenever one contact was unpolarized and the other had finite polarization.  The negative signals were reproducible and observed in all channels measured, for all cooldowns, indicating that they reflect an intrinsic phenomenon.

The negative signal can be explained by different conductivities for spin-up and spin-down carriers $\sigu > \sigd$, see Fig.~2(b).  If $P_{inj}=1$, spin-up electrons  accumulate above the injector, giving $\mu_\uparrow>\mu_\downarrow$ as in Fig.~2(a). But the spin-up current diffusing {\em towards} the floating reservoir must be balanced by a spin-down current {\em from} the floating reservoir that is impeded by a lower spin-down conductivity.    In order to maintain zero net charge current on the right side, $\mu(R)$ must equilibrate at a chemical potential closer to $\mu_\uparrow$ than to $\mu_\downarrow$.  This makes $\vnl=\mu_{av}(x_{det})-\mu(R)$ negative for \{1,0\} (Fig.~2(b)).

A similar argument explains the negative $\vnl$ for \{0,1\} (Fig.2(b)).  When the injected current is unpolarized, equal currents of spin-up and spin-down must flow to the drain, but the lower conductivity for spin-down requires a larger $\mu_\downarrow$ compared to $\mu_\uparrow$ in the channel.  The floating reservoir equilibrates at a potential somewhere between $\mu_\uparrow(0)$ and $\mu_\downarrow(0)$, so the voltage measured at \{0,1\} is negative ($\mu(R) > \mu_\uparrow(x_{det})$). The negative signals at \{0,1\} and \{1,0\} are connected by the Onsager relation: when current and voltage probes were switched and the magnetic field was reversed, \{0,1\} and \{1,0\} signals were found to be identical, as expected.

Proof that the negative signals at \{0,1\} and \{1,0\} are due to spin accumulation can be found in their magnetic field dependence: both disappear due to spin-orbit-mediated spin relaxation.  Typically the spin relaxation length is greater than the injector-detector separation, $x_{det}$, ensuring a measurable spin signal at the detector.\cite{Frolov}  But the relaxation length collapses when in-plane magnetic fields of the proper magnitude are applied perpendicular to the channel direction, that is, along $\hat{y}$.  The mechanism for the enhanced relaxation is ballistic spin resonance (BSR), which occurs when $g\mu_B B_y/h=2\tau_c^{-1}$, where $g$ is the Lande g-factor, $h$ is the Planck constant, and $\tau_c$ is the time for electrons to cross the channel.\cite{Nature}

Figure 3(a) shows the $B_y$-dependence of positive $\vnl$ (\{1,1\}) and negative $\vnl$ (\{0,1\}) for a micron-wide channel.  The collapse in the spin relaxation length due to BSR causes a collapse in $\vnl$ near $B_y=6T$ for both \{1,1\} and \{0,1\}, indicating that both positive and negative signals arise from spin polarization.  Both positive and negative signals disappear at low field because the measurement requires a spin-sensitive detector QPC, and QPC polarization turns on only at fields of a few Tesla.  (The residual voltages at zero field are signatures of the Peltier effect and not of spin polarization.\cite{Frolov})

\begin{figure}[ht]
  \begin{center}
  \includegraphics[scale=1]{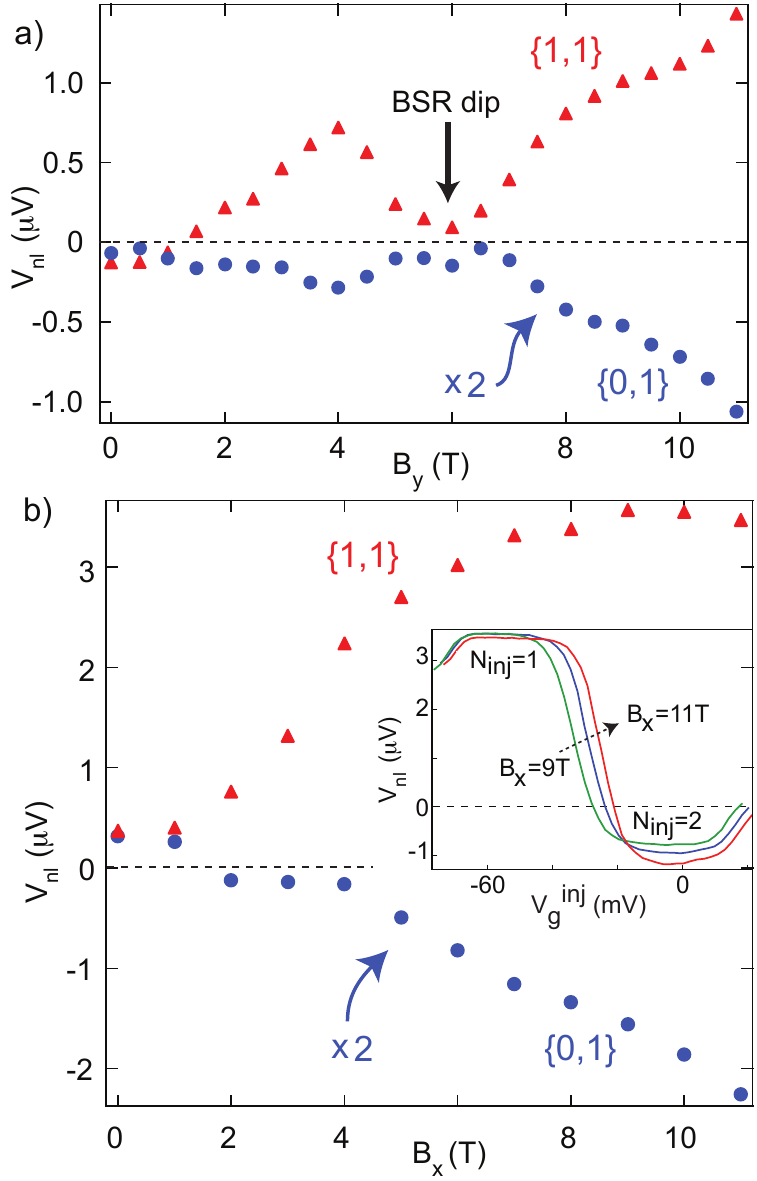}
  \end{center}
  \caption{ \textbf{(a, b)} Positive \{1, 1\} and negative \{0, 1\} nonlocal signals vs. $B_y$ and $B_x$.  Negative signals are multiplied by 2 for clarity. The negative signal at \{1,0\} displayed a similar field dependence, but is not shown here.  Inset: Example of data from which points in (b) were extracted, showing evolution of nonlocal signal for both injector settings.   Data in panels (a) and (b) are from different devices.}
  \label{fig:bsr}
\end{figure}

In order to understand the evolution of the negative signal for fields along the channel axis $\hat{x}$ direction (Fig.~3(b)), where no BSR is expected, we consider a simple model describing conductivities for spin-up and spin-down carriers in an in-plane field. When the Zeeman splitting is comparable to the Fermi energy, the two populations have significantly different densities, $n_{\uparrow(\downarrow)}$, and therefore  different Fermi velocities: $v^F_{\uparrow(\downarrow)}=\sqrt{2({E^0_F\pm g^*\mu_BB/2)/m^*}}=\sqrt{\frac{4n_{\uparrow(\downarrow)}}{\varrho m^*}}$, where $E^0_F=(n_\uparrow+n_\downarrow)/\varrho$ represents the Fermi energy at zero magnetic field with spin-degenerate 2D density of states $\varrho=m^*/\pi\hbar^2$, $m^*$ is the effective electron mass and $g^*$ is the effective g-factor. For a given mean free path, $\lambda_e$, the difference in $v_F$ leads to different conductivities,
\begin{equation}\label{spincond}
\sigma_{\uparrow(\downarrow)}=\varrho e^2\lambda_e\frac{v^F_{\uparrow(\downarrow)}}{4}= e^2\lambda_e\sqrt{\frac{\varrho n_{\uparrow(\downarrow)}}{4m^*}}.
\end{equation}

The quantitative relation between spin-dependent conductivities and $\vnl$ can be extracted from a 1D diffusion equation that includes spin relaxation, $\partial^2\vnl/\partial x^2=\vnl/\lambda_s^2$.\cite{Jedema}  Defining independent conductivities $\sigma_D$, $\sigma_M$, and $\sigma_R$ for the drain, middle, and reservoir segments (Fig.~1), and matching boundary conditions between drain/middle and middle/reservoir segments with  $\mu_\uparrow=\mu_\downarrow$ enforced at both ends of the channel, one obtains:
\begin{equation}\label{Vnlequ}
    \Vnl(\Pinj,\Pdet)=\Pinjhat\times\Pdethat\times \Gamma,
\end{equation}
where $\hat{P}$ is an {\em effective} contact polarization that includes the effect of spin-resolved conductivities in the bulk:
\begin{equation}\label{Peff}
    \hat{P}_{\mathrm{inj (det)}}=P_{\mathrm{inj(det)}}-\frac{\sigma_{\mathrm{D(R)}\uparrow}-\sigma_{\mathrm{D(R)}\downarrow}}{\sigma_{\mathrm{D(R)}\uparrow}+\sigma_{\mathrm{D(R)}\downarrow}}.
    \end{equation}
$\Gamma$ depends on spin-dependent conductivities in all three channel segments, the spin relaxation length, geometrical parameters and injector current, but {\em not} on the QPC polarizations $\Pinj$ and $\Pdet$.    

Eqs.~(\ref{spincond}-\ref{Peff}) explain the magnetic field dependences for fields along $\hat{x}$ (Fig.~3(b)).  Both positive and negative signals increased from zero for $B_x<5$T, reflecting the increase in QPC polarization. The positive signal saturated when QPC polarization reached 100\%, but the negative signal continued to grow with field because $g^*\mu_B B_x$ was less than $\epsilon_F$ throughout the accessible field range.    These equations also explain the locations of positive and negative signal in Fig.~2(c), taking into account that $\sigu>\sigd>0$.  The effective polarization $\hat{P}$ is positive when QPCs are fully polarized ($P=1$), but $\hat{P}$ is negative for $P=0$, so $V_{nl}$ is negative for \{1,0\} and \{0,1\}.

%The solid lines in Fig.~3(b) are a fit to the positive and negative signal data, using total density $n_{\uparrow}+n_\downarrow$ extracted from Shubnikov-deHaas oscillation periodicity at $B_x=0$.  The fit includes three free parameters, constrained to be the same for the positive and negative signal data: energy broadening in the QPC\cite{Frolov}, spin susceptibility $m^*g^*$, and the overall scaling factor $\Gamma$ from Eq.~\ref{Vnlequ}.   The spin susceptibility extracted with this fit was significantly enhanced above the bare susceptibility, $m^*g^*=(4\pm0.4)\times m_bg_b$, using bare effective mass $m_b=0.067m_e$ and bare g-factor $|g|=0.44$.  The value is consistent with previous measurements of field-dependent effective mass and density-dependent g-factormade using tilted field techniques.\cite{Tutuc3, Tan}

\begin{figure}[t]
  \includegraphics[scale=1.0]{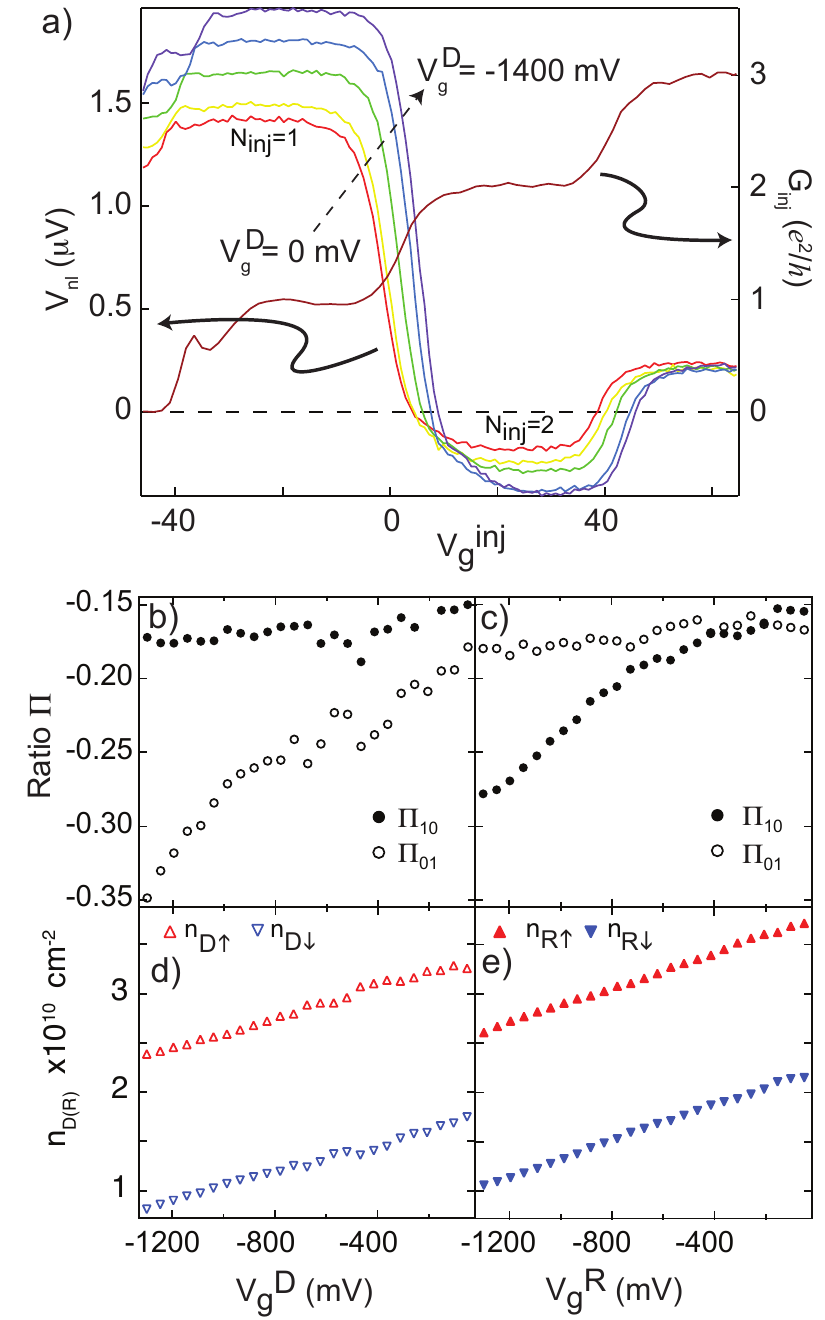}
  \caption{\textbf{(a)} Left axis: Channel gate $\Vl$ evolution of the nonlocal signal sweeping the injector across the first two plateaus (detector polarized).  Right axis: conductance of the injector QPC. \textbf{(b, c)} Ratios $\Pi_{01}$ and $\Pi_{10}$ calculated from positive and negative extrema of nonlocal signal traces like those in (a). \textbf{(d, e)} Spin-resolved densities in drain (d) and reservoir (e) segments.  Data in (d) extracted from $\Pi_{01}$ in (b); data in (e) extracted from $\Pi_{10}$ in (c).  ($B_x=10$T in all panels.)}
  \label{}
\end{figure}

We now turn our attention to the high magnetic field regime where the QPC transmission can be fully polarized.  Figure 4 explores the effect of changing the voltages on the channel-defining gates, $\Vl$ or $\Vr$, at  $B_x$=10T.  When $\Vl$ is made more negative, the drain segment of the channel is narrowed and the electron density $n_\mathrm{D}$ is reduced.  Similarly, more negative values of $\Vr$ lead to a narrower reservoir segment of the channel, and to smaller $n_\mathrm{R}$. (Note that the changes in $\Vl$ and $\Vr$ required to cause a significant change in the density are much larger than the changes in $\Vinj$ and $\Vdet$ required to go between \{0,1\}, \{1,0\} and \{1,1\}.)

Negative voltages on $\Vl$ and $\Vr$ affect the nonlocal signal in two important ways.  First, the total conductance (spin-up and spin-down together) is reduced as a result of the narrower channel geometry and shorter mean free path, leading to a trivial increase in the magnitude of nonlocal voltage when the current is held fixed (positive voltages become more positive, negative voltages become more negative, Fig.~4(a)).  Second, lower density gives a smaller Fermi energy, therefore a larger fractional difference between majority and minority spin populations and conductivities at finite field (Eq.~1).  The negative signal is much more strongly influenced by $\sigma_{\uparrow}/\sigma_\downarrow$ than the positive signal is (Eq.~3): in Fig.~4(a) the positive signal grows by only $\sim30$\%, whereas the negative signal grows by more than a factor of two.

Because $\Gamma$ from Eq.~3 does not depend on the QPC polarizations explicitly, the ratio between negative and positive signals directly determines the ratio in the conductivities, $\sigma_{\uparrow}/\sigma_\downarrow$, for each value of gate voltage (Fig.~4(b),(c)):
\begin{equation}\label{ratio}
    \Pi_{01 (10)}=\frac{1}{2}\left(1-\frac{\sigma_{\mathrm{D(R)}\uparrow} }{ \sigma_{\mathrm{D(R)}\downarrow} }\right).
\end{equation}  where $ \Pi_{01}\equiv\Vnl(0,1)/\Vnl(1,1)$ and $ \Pi_{10}\equiv\Vnl(1,0)/\Vnl(1,1)$.  The difference in the effects of drain and reservoir conductivities predicted in Eq.~4 provides an additional test for the proposed origin for the negative signal.  According to Eq.~4, $ \Pi_{01}$ depends exclusively on drain conductivities, $\sdu$ and $\sdd$, while $ \Pi_{10}$ depends exclusively on reservoir conductivities $\sru$ and $\srd$.    This distinction is clearly visible in Figs.~4(b),4(c): the drain gate $\Vl$ affects  $ \Pi_{01}$ strongly while $ \Pi_{10}$ is essentially unchanged;  the reservoir gate $\Vr$ affects  $ \Pi_{10}$ strongly while $ \Pi_{01}$ is unchanged.

The ratio $\sigma_{\uparrow}/\sigma_\downarrow$ leads directly to the ratio between spin-resolved carrier densities, $n_{\uparrow}/n_\downarrow$, if one assumes a simple scattering model in which the mean free path is independent of spin direction even at finite polarization.  The total charge densities, $n_{\uparrow}+n_\downarrow$, in the drain and reservoir segments were measured for each gate voltage using Shubnikov-de Haas periodicity at zero in-plane field.  Together, $n_{\uparrow}/n_\downarrow$ and $n_{\uparrow}+n_\downarrow$  fix the values for both $n_{\uparrow}$ and $n_\downarrow$ at 10T, plotted in Figs.~4(d),(e).   The difference $n_{\uparrow}-n_\downarrow=1.6\pm0.1\times10^{10}$cm$^{-2}$ does not change with gate voltage, indicating an enhanced spin susceptibility 4.5 times the bare value that does not depend strongly on density within the range $3\times10^{10}-6\times10^{10}$  cm$^{-2}$.  This enhancement is consistent with values reported in Refs.~\onlinecite{Tutuc3} and \onlinecite{Tan}, which included both field-induced enhancement of the effective mass and exchange enhancement of the effective g-factor.

The significant difference between populations of spin-up and -down carriers that is reflected in Figs.~4(d),(e) suggests that the assumption of equal mean free paths for both spins is a poor approximation, especially for the high levels of polarization reached at very negative gate voltage.  We are not aware of theoretical calculations that predict spin-resolved scattering rates in a partially polarized 2DEG.  At a qualitative level one would expect a longer mean free path for majority carriers, $\lambda_{\uparrow}>\lambda_{\downarrow}$, and that the ratio $\lambda_{\uparrow}/\lambda_{\downarrow}$ would increase with the ratio of densities $n_{\uparrow}/n_{\downarrow}$.

This trend would tend to decrease the susceptibility that is extracted from the data, and the decrease would be greatest when $\lambda_{\uparrow}/\lambda_{\downarrow}$ was large.  As a result, the data  in Figs.~4(d,e) would indicate that the susceptibility {\em grows} with density in the range $3\times10^{10}-6\times10^{10}$  cm$^{-2}$ even at fixed field---a result that has not been predicted in the literature to our knowledge. More careful quantitative analysis will require calculations that consider the effects of small-angle scattering  (the type of scattering expected to be dominant in high-mobility heterostructures) in a partially-polarized 2DEG.

{\bf Acknowledgements:}  Work at UBC supported by NSERC, CFI, and CIFAR.  W.W. acknowledges financial support by the Deutsche Forschungsgemeinschaft (DFG) in the framework of the program  ``Halbleiter-Spintronik'' (SPP 1285).


\begin{thebibliography}{99}
\bibliographystyle{alpha}

\bibitem{Zutic} I. Zutic, J. Fabian, S. Das Sarma, Rev. Mod. Phys. {\bf 76}, 323 (2004).
\bibitem{Gao} X. P. A Gao {\em et al.}, Physical Review B {\bf 73} 241315(R) (2006).
\bibitem{Piot} B. A. Piot {\em et al.},  Physical Review B {\bf 80} 115337 (2009).
\bibitem{Tutuc2} E. Tutuc {\em et al.}, Physical Review Letters {\bf 88}, 036805 (2002).
\bibitem{Gold} V.T. Dolgopolov, A. Gold, JETP Lett. {\bf 71}, 27 (2000).
\bibitem{Liang} C. Liang {\em et al.}, Physica E {\bf 18}, 141 (2003).
\bibitem{Tan} Y.-W. Tan {\em et al.}, Physical Review B {\bf 73}, 045334 (2006).
\bibitem{PotokPRL02} R. M. Potok {\em et al.}, Physical Review Letters {\bf 89}, 266602 (2002).
\bibitem{Johnson} M. Johnson and R. H. Silsbee, Physical Review Letters {\bf 55}, 1790 (1985).
\bibitem{Frolov} S. M. Frolov {\em et al}, Physical Review Letters {\bf 102}, 116802 (2009).
\bibitem{Jedema} F. J. Jedema et al, Journal of Superconductivity: Incorporating Novel Magnetism, Vol. 15, page 27 (2002).
\bibitem{PotokPRL03} R. M. Potok {\em et al.}, Physical Review Letters {\bf 91},  016802 (2003).
\bibitem{HansonPRB04} R. Hanson {\em et al.}, Physical Review B  {\bf 70}, 241304(R) (2004).
\bibitem{Wees} B. J. van Wees {\em et al}, Physical Review Letters {\bf 60}, 848 (1988).
\bibitem{Wharam} D. A. Wharam {\em et al}, Journal of Phys. C-Solid State Phys. {\bf 21}, L209 (1988).
\bibitem{Nature} S. M. Frolov, {\em et al.}, Nature {\bf 458}, 868 (2009).
\bibitem{Tutuc3} E. Tutuc {\em et al.}, Physical Review B {\bf 67}, 241309(R) (2003).


\end{thebibliography}
\end{document}